\documentclass[aps,
twocolumn,
showpacs,preprintnumbers,amsmath,amssymb,prb,superscriptaddress]{revtex4-1}
\usepackage{graphicx}
\usepackage{amssymb}
\usepackage{amsmath}
\usepackage{enumitem}
\usepackage{color}
\usepackage{textcomp}
\usepackage[symbol]{footmisc}

\begin{document}

\title{Fabrication of comb-drive actuators\\ for straining nanostructured suspended graphene}

\author{M.~Goldsche}
\affiliation{JARA-FIT and 2nd Institute of Physics, RWTH Aachen University, 52074 Aachen, Germany}
\affiliation{Peter Gr\"{u}nberg Institute (PGI-8/9), Forschungszentrum J\"{u}lich, 52425 J\"{u}lich, Germany}
\author{G.~J.~Verbiest}
\affiliation{JARA-FIT and 2nd Institute of Physics, RWTH Aachen University, 52074 Aachen, Germany}
\author{T.~Khodkov}
\email{Current address: ICFO, Barcelona, Spain}
\affiliation{JARA-FIT and 2nd Institute of Physics, RWTH Aachen University, 52074 Aachen, Germany}
\affiliation{Peter Gr\"{u}nberg Institute (PGI-8/9), Forschungszentrum J\"{u}lich, 52425 J\"{u}lich, Germany}
\author{J.~Sonntag}
\affiliation{JARA-FIT and 2nd Institute of Physics, RWTH Aachen University, 52074 Aachen, Germany}
\affiliation{Peter Gr\"{u}nberg Institute (PGI-8/9), Forschungszentrum J\"{u}lich, 52425 J\"{u}lich, Germany}
\author{N.~von~den~Driesch}
\affiliation{Peter Gr\"{u}nberg Institute (PGI-8/9), Forschungszentrum J\"{u}lich, 52425 J\"{u}lich, Germany}
\author{D.~Buca}
\affiliation{Peter Gr\"{u}nberg Institute (PGI-8/9), Forschungszentrum J\"{u}lich, 52425 J\"{u}lich, Germany}
\author{C.~Stampfer}
\affiliation{JARA-FIT and 2nd Institute of Physics, RWTH Aachen University, 52074 Aachen, Germany}
\affiliation{Peter Gr\"{u}nberg Institute (PGI-8/9), Forschungszentrum J\"{u}lich, 52425 J\"{u}lich, Germany}

\begin{abstract}
We report on the fabrication and characterization of an optimized comb-drive actuator design for strain-dependent transport measurements on suspended graphene. We fabricate devices from highly p-doped silicon using deep reactive ion etching with a chromium mask. Crucially, we implement a gold layer to reduce the device resistance from $\approx51.6$ k$\mathrm{\Omega}$ to $\approx236$ $\mathrm{\Omega}$ at room temperature in order to allow for strain-dependent transport measurements. The graphene is integrated by mechanically transferring it directly onto the actuator using a polymethylmethacrylate membrane. Importantly, the integrated graphene can be nanostructured afterwards to optimize device functionality. The minimum feature size of the structured suspended graphene is 30~nm, which allows for interesting device concepts such as mechanically-tunable nanoconstrictions. Finally, we characterize the fabricated devices by measuring the Raman spectrum as well as the a mechanical resonance frequency of an integrated graphene sheet for different strain values.
\end{abstract}

\maketitle

\section{Introduction\label{S1}}

Graphene offers unique combinations of electrical and mechanical properties \cite{Novoselov666}, which allows for flexible electronics \cite{kim2015} as well as high-frequency resonators \cite{chen2009performance,bunch2007,eichler2011}, sensors \cite{bogue2014,verbiest2018}, filters \cite{jiang2016}, mixers \cite{zhou2013,zhou2015,heath2017} and amplifiers \cite{zhou2013,zhou2015,heath2017}.
The application of strain alters the mechanical properties, e.g. the resonance frequency of suspended graphene devices \cite{chen2009performance,bunch2007}, but more remarkably also the electronic properties.
Strain induces a pseudo-vector similar to that of a real magnetic field \cite{pereira2009,guinea2010,verbiest2015}.
Control over strain fields \cite{goldsche2018} may lead to fully strain engineered devices such as valley-filters \cite{pereira2009} as well as piezoelectric \cite{smith2016} and superconducting \cite{uchoa2013} devices.
However, the experimental realization of such devices has proven to be a major challenge.
In commonly used methods, such as pulling on suspended flakes with a bottom-electrostatic gate or bending a flexible substrate, the obtained strain fields are intrinsically linked to either the electronic tuning of the charge carrier density \cite{chen2009performance,bunch2007,song2011,zhang2014,guan2015,nicholl2015,bao2012}or to the properties of the substrate \cite{mohiuddin2009uniaxial,yoon2011strain,huang2010,zhang2014,guan2015}.
Here we report on the fabrication and characterization of surface micromachined silicon-based comb-drive (CD) actuators to strain suspended graphene.
The devices are low-temperature compatible and allow to decouple the obtained strain fields from the electronic tuning of the charge carrier density and from the properties of the substrate.
In addition, we implement a gold layer on the CD actuators to reduce the contact resistance to the suspended graphene in order to enable transport measurements.
Finally, we introduce a technique to pattern suspended graphene with a minimum feature size of 30~nm after its mechanical transfer onto the CD actuators to optimize device functionality and to enable mechanically-tunable suspended nanoconstriction devices and potentially suspended quantum dots.

\begin{figure*}[!t]
    \centering
    \includegraphics[width=\textwidth]{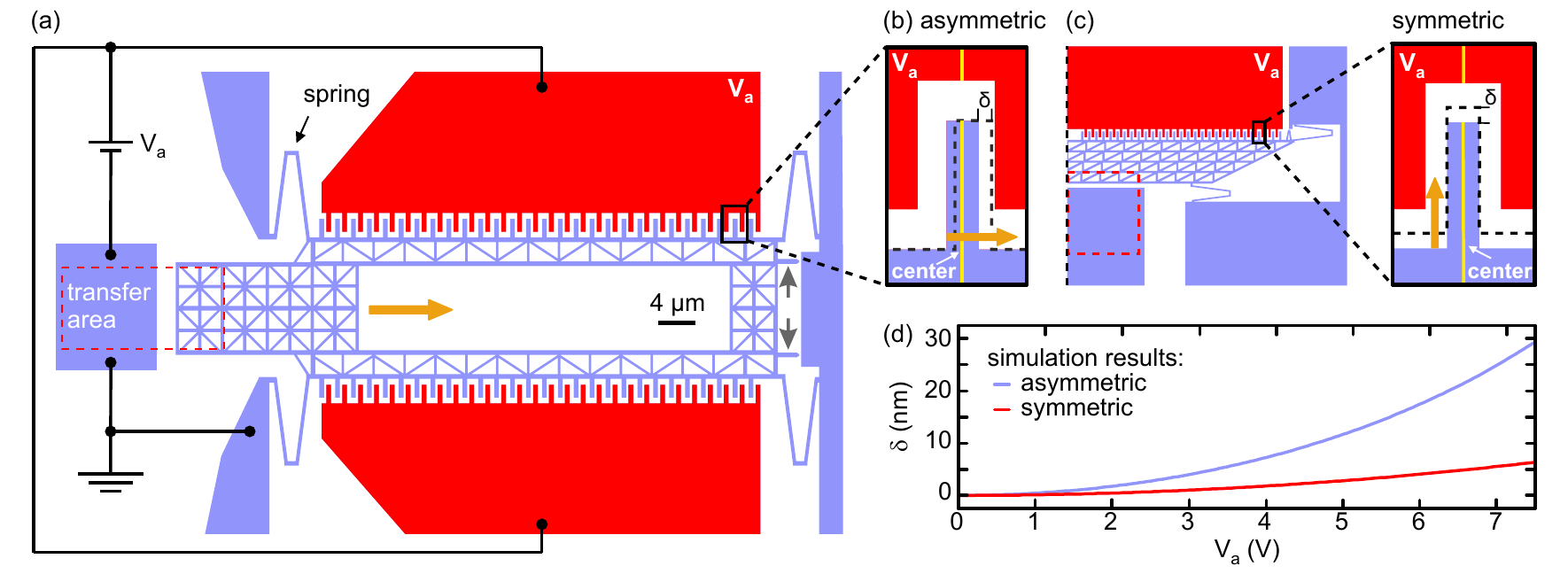}
	\caption{(a) Schematic illustration of the actuator design considered in this work.
             The gray arrows indicate the stoppers (see text).
             We place the fingers on purpose asymmetric (b) with respect to the center, which is in contrast with the common symmetric placement (c) of the fingers.
             Note that we only show half of the symmetric design in the left panel in (c) as indicated by the vertical black dashed line.
             In (a) and (c), the red box indicates the area where we transfer the graphene and the orange arrows indicate the direction of motion.
             (d) A finite element simulation shows that design (a) has up to 5 times larger displacement $\delta$ than design (c) for the same actuator potential $V_\text{a}$, finger overlap area, and total number of fingers.}
	\label{fig1}
\end{figure*}

A CD actuator consist of two interdigitated combs (see figure~\ref{fig1}(a)) of which one is fixed to a substrate and one that is held suspended by springs.
By applying a potential difference between the fingers of the combs, an electrostatic force arises due to their capacitive coupling resulting in a displacement of the suspended comb.
The important design parameters for CD actuators are the overlap area of the fingers \cite{legtenberg1996}, the shape of the fingers \cite{jensen2003shape,goldsche2014} and the distance between them \cite{legtenberg1996}.
To maximize the capacitive coupling, a high overlap area and small finger separation is desired.
Consequently, the comb structures have high aspect ratios: the thickness of the fingers is much larger than their separation.

CD actuators are usually surface micromachined from polycrystalline silicon \cite{judy2001,sasaki2003,2005poymumps} that is grown on a sacrificial layer, e.g silicon oxide.
The comb structures are patterned via optical lithography \cite{lorenz1998} and then carved into the polycrystalline silicon by reactive ion etching (RIE) \cite{rangelow2003}.
Finally, one of the combs is released from the substrate by removing the sacrificial silicon oxide layer with (buffered) hydrofluoric acid (HF) \cite{Proksche1992,spierings1993hf} or hydrofluoric vapor \cite{helms1992hfvapor}.
Standard optical lithography limits the minimal feature size, e.g. the finger width or the distance between the fingers, to around 0.5~$\mu$m.
Concomitantly, the low capacitive coupling requires high voltages to reach the maximum displacement of the CD actuator, e.g. the design in Ref. \onlinecite{Li2012Comb} needs 400 V.
When reducing the minimal feature size into the nanometer regime with electron beam lithography, the scallops resulting from the RIE process on the vertical surface of the CD fingers become increasingly important and decreases the reliability of the actuator \cite{gaither2013processeffects}.
Along the same line, polycrystalline silicon actuators typically show a lower reliability than monocrystalline silicon. 
We therefore fabricate our low-temperature compatible CD actuators based on highly p-doped monocrystalline silicon using electron beam lithography.

The paper is organized as follows.
We first introduce our actuator design (Sect.~\ref{S2}).
This is then followed by a detailed description of the fabrication process in Sect.~\ref{S3}.
This includes a subsection for the substrate preparation, the deep reactive ion etch process, the integration of a thin Au film, the transfer of graphene, the patterning of suspended graphene, and the release of the actuator from the substrate.
Finally, we characterize the fabricated actuators in Sect.~\ref{S4} by measuring the induced strain in the graphene with the help of spatially-resolved scanning confocal Raman spectroscopy \cite{graf2007,neumann2015b,goldsche2018} and by measuring its tunable mechanical resonance frequency with an amplitude modulated down-mixing technique \cite{verbiest2016}.

\section{Design\label{S2}}

To ensure low operating voltages, we reduce the minimal feature size down to 60~nm.
A schematic illustration of our actuator design is depicted in figures~\ref{fig1}(a-b).
Our CD design differs from commonly used CD designs \cite{tang1989laterally, xie2002vertical} (see e.g. figure~\ref{fig1}(c)) at two points.
We change the direction of motion (indicated with the orange arrows in figure~\ref{fig1}) in comparison to common actuators by designing the interdigitated fingers on purpose asymmetric (figure~\ref{fig1}(b)), i.e. unequal distances between the fingers.
In figure~\ref{fig1}(c), we show the standard symmetric placement of the fingers and resulting direction of motion for comparison.
To prevent the suspended comb from moving in the direction indicated by the orange arrow in figure~\ref{fig1}(c), we apply interdigitated fingers to both sides of the suspended comb.

We compared the performance of our CD design in figure~\ref{fig1}(a) with the symmetric design in figure~\ref{fig1}(c) using a finite element simulation \cite{comsol}.
In these simulations, the finger overlap area and total number of fingers were for both CD designs the same.
We computed the actuator displacement $\delta$ as a function of applied actuator potential $V_\text{a}$.
Figure~\ref{fig1}(d) shows that our CD designs reaches up to 5 times more displacement than the symmetric design for the same $V_\text{a}$.
In addition, we increased the length of the actuator along the direction of motion to decrease the rotation resulting from a non-perfectly transferred graphene flake (see Sect. 3.5).
Note that the maximum displacement of the actuator is for the asymmetric design lower than for the symmetric design.
However, the maximum displacement translates to over 4\% of strain for a graphene flake with a typical suspended length of 2~\textmu m, which exceeds the experimentally observed breaking strength in doubly clamped suspended graphene \cite{goldsche2018}.
As such the travel range is not the limiting factor, especially considering that the design of the actuator is easily adjustable to allow for higher maximum displacements.
Similar to common actuators, the suspended comb body consist of a grated structure to ensure release from the substrate, and has integrated stoppers (see gray arrows in figure~\ref{fig1}(a)) to limit its displacement and thereby prevent an electrical connection between the interdigitated fingers.
Finally, we point out the possibility to integrate an electrostatic gate in the design underneath the transfer area (dashed red boxes in figures~\ref{fig1}(a) and \ref{fig1}(c)) to additionally tune the charge carrier density in the graphene.

\section{Fabrication\label{S3}}

\subsection{Substrate preparation\label{S31}}
We start with a commercially available (100) silicon-on-insulator (SOI) wafer (G8P-214-01, Unibond$^\text{TM}$ wafers, Soitec) with a top layer of 200~nm and a buried oxide thickness of 1~\textmu m.
First, we use dry thermal oxidation at 1050~$^{\circ}$C to obtain a 50 nm thick silicon oxide (SiO$_2$) layer.
Second, we remove the SiO$_2$ layer with HF chemistry. By repeating these steps three times, we are left with a SOI wafer with a top Si layer of 50~nm.
Third, we use an industry-compatible reduced-pressure chemical vapor deposition (CVD) system for the epitaxial growth of a 2~\textmu m thick highly p-doped silicon layer on top of the SOI wafers.
Prior growth, the native oxide was removed ex-situ using HF vapor chemistry.
A subsequent in-situ hydrogen bake ensures a contaminant-free, pristine silicon surface.
For the epitaxial growth, we use disilane (Si$_2$H$_6$) and diborane (B$_2$H$_6$) as a precursor at a temperature of 850~$^{\circ}$C to ensure elevated growth rates and a high boron dopant concentration of about $1.5 \times 10^{19}$~cm$^{-3}$.
This yields a sheet resistance of $\approx$ 43.7~$\Omega$/$\Box$  and ensures full functionality of the device at cryogenic temperatures (T $\approx$ 10 mK ) \cite{verbiest2016}.

\subsection{Deep reactive ion etching process\label{S32}}
\begin{figure*}
	\makebox[\textwidth][c]{\includegraphics[width=\textwidth]{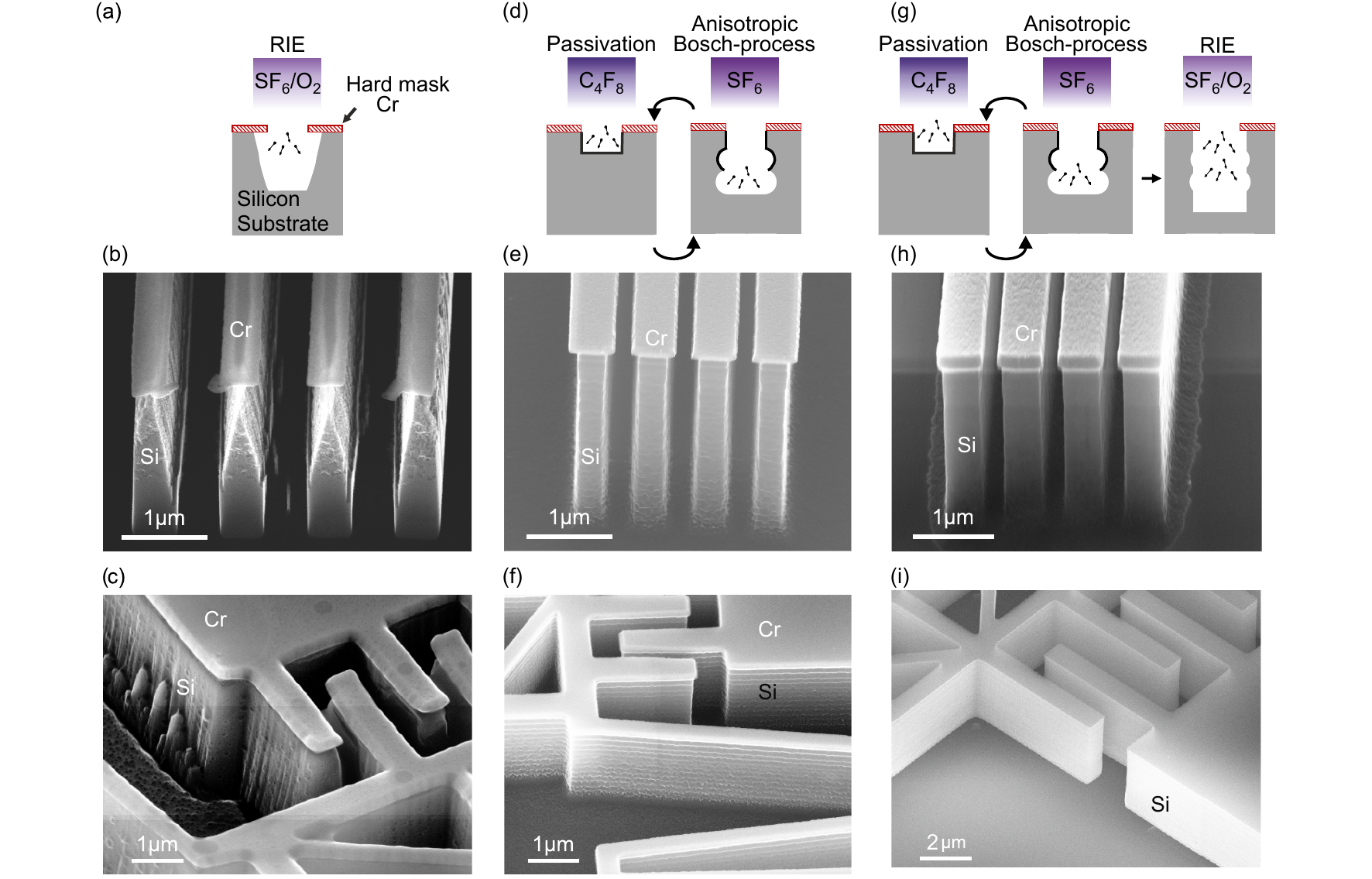}}%
	\caption{Schematic illustrations and scanning electron microscope images of different etch processes used to reach a depth of 2~\textmu m: (a-c) a standard RIE process using SF$_6$/O$_2$ without ICP generates an unacceptably large undercut underneath the Cr mask, (d-f) a Bosch DRIE process achieves aspect ratios of up to 10:1, but also generates significantly large scallops, (g-i) a standard optimized RIE process using an additional SF$_6$/O$_2$ etch step after the DRIE process depicted in panel (d) reduces the scallops to below 5~nm (see inset in panel (h)), which is required for optimal CD actuator performance.}
	\label{fig2}
\end{figure*}
The 2~\textmu m thick p-doped silicon layer in combination with a minimal feature size of the fingers of 200~nm is demanding on the deep reactive ion etching (DRIE) that is widely used in surface machining and consequently for structuring CD actuators.
The minimal feature size requires electron beam lithography, as it is below the resolution of our standard optical lithography.
Compared to standard optical lithography, electron beam lithography offers the flexibility and precision to rapidly prototype and optimize the actuator design.
However, electron beam resists such as polymethylmethacrylate (PMMA) are generally not suitable for patterning CD actuators, as their DRIE etch rates are comparable to that of the substrate \cite{chinn1981RIE,wuest2005fab}.
We circumvent this via the deposition of a chromium (Cr) hard mask, which has a negligible edge rate in reactive ion etching processes \cite{shearn2010}.
Chromium is a standard mask in III-V semiconductor processing which is known to have a small line edge roughness \cite{shearn2010} and is easily removed with a mixture of perchloric acid (HClO$_4$), and ceric ammoniumnitrate (NH$_4$)$_2$[Ce(NO$_3$)$_6$].
The Cr mask is patterned in a resist stack of copolymer (AR617.04, 250 nm) and PMMA (AR679.04, 270 nm) using electron beam lithography.
Most importantly, the Cr mask allows for the implementation of standard DRIE processes.
Nevertheless, the large aspect ratio of 10:1 is problematic, as any deviation from the ideal etching behavior will directly affect the device functionality, relying on the rather homogeneous capacitance between the individual fingers.
Note that the issue is even more problematic for the stoppers (gray arrows in figure~\ref{fig1}a), as their separation from the fixed anchor is just 60~nm and thus requires an aspect ratio of 33:1.
\begin{figure}[!b]
	\includegraphics[width=84mm]{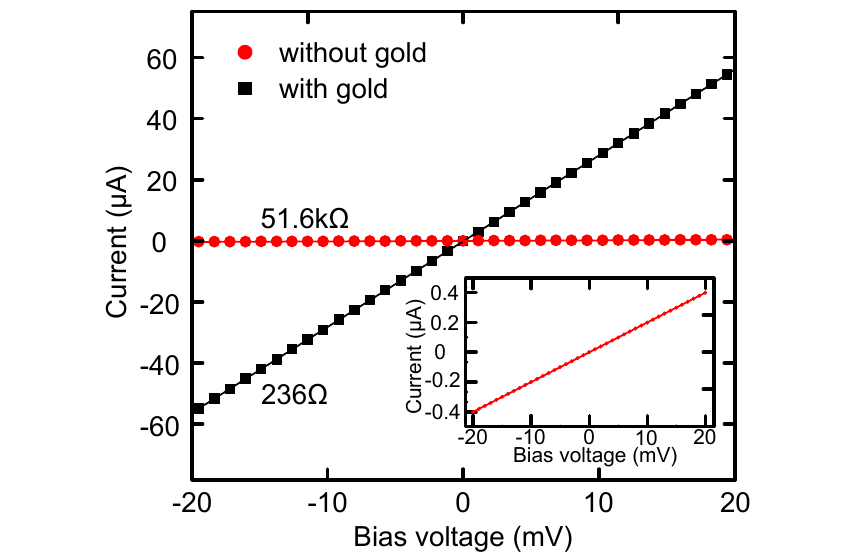} 
	\caption{The current through the actuator as a function of applied bias voltage for a device at room temperature without gold (red) and with gold layer (black). The inset provides a zoom-in of the measurement without the gold layer. The gold lowers the resistance through the device from 51.6~k$\mathrm{\Omega}$ to 236~$\mathrm{\Omega}$.}
	\label{fig3}
\end{figure}
\begin{figure*}[hbt!]
	\makebox[\textwidth][c]{\includegraphics[width=\textwidth]{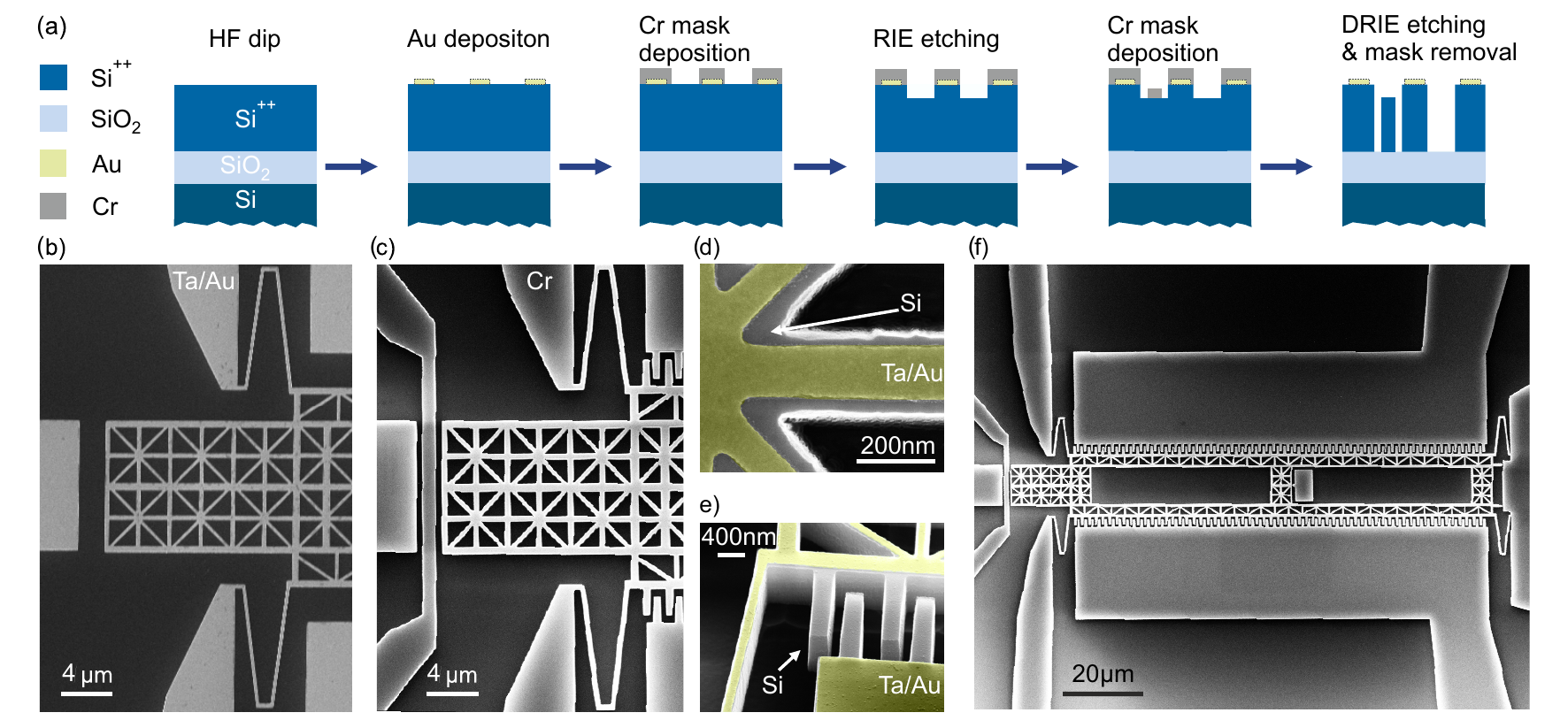}}%
	\caption{(a) A schematic illustration of the process steps taken to integrate a metal onto the actuator and to fabricate an electrostatic gate underneath the transfer region. The scanning electron microscope images (b-f) show different steps of the fabrication process: (b) After HF dip and Au deposition, (c) after Cr mask deposition, (d) and (e) show a top and side view of the implemented metal after DRIE and Cr mask removal. (f) Top view of the final device.}	\vspace{-10pt}
	\label{fig4}
\end{figure*}

Figure~\ref{fig2}(a) shows a schematic illustration of a standard reactive ion etch with a mixture of SF$_6$/O$_2$ without inductively coupled plasma (ICP).
As the tilted scanning electron microscope (SEM) images in figures~\ref{fig2}(b-c) show, this results in a notching structure underneath the Cr mask.
The notching structure is caused by a significant sideways etching rate of the silicon due to sidewall charging \cite{chen2012lecture,welch2011nanoscale}.
Therefore, we use a Bosch DRIE process \cite{lui2013cryetch}.
A schematic illustration of this process is depicted in figure~\ref{fig2}(d).
We use C$_4$F$_8$ with a radio frequency (RF) power of 24~W and an ICP power of 300~W for 10~s to generate a passivation layer.
This is followed by SF$_6$ with a RF power of 12~W and an ICP power of 200~W for 4~s to isotropically etch the substrate.
We repeat these steps 20~times to etch 2~\textmu m deep.
This Bosch DRIE process does not generate a notching structure underneath the Cr mask (see the tilted SEM image in figure~\ref{fig2}(e)).
However, as the tilted SEM figure~\ref{fig2}(f) shows, this Bosch DRIE process generates scallops and thus results in a higher sidewall roughness.
As this has been identified as one of the main sources of the reduced fracture strength \cite{gaither2013processeffects}, we implemented an additional etching step with a mixture of SF$_6$/O$_2$ that has a RF power of 60~W for 60~s after the final SF$_6$ step (see illustration in figure~\ref{fig2}(g)).
As figure~\ref{fig2}(h-i) shows, we do not see evidence that the additional etching with SF$_6$/O$_2$ causes a notching structure underneath the Cr mask (see figures~\ref{fig2}(a-c)).
However, the additional etching step dramatically reduces the scallop depth and thus the sidewall roughness to below 5~nm (see inset in figure~\ref{fig2}(h)).
The width of the fingers of the structure in figure~\ref{fig2}(h) is only $\approx$130~nm smaller than the Cr mask, which corresponds to an average undercut of $\approx$65~nm.
Finally, we point out that the width of the fingers in the devices reported in Ref. \onlinecite{goldsche2018} is only $\approx$30~nm smaller than the Cr mask, which corresponds to an average undercut of $\approx$15~nm.

\subsection{Integration of thin Au film\label{S33}}
To study charge carrier transport through graphene, the high resistance of the actuator ($\approx$50~k$\Omega$ at room temperature (red data points in figure~\ref{fig3}) and $\approx$1~M$\Omega$ at 2.2~K \cite{verbiest2016}) poses a major challenge.
The high device resistance diminishes any resistance change of the graphene in two-terminal transport measurements.
We solve this problem by depositing a thin metal layer on the actuator before the deposition of the Cr mask needed for the actuator fabrication (see above).
It is not possible to do this after the actuator fabrication due to the large thickness of the fingers (2~\textmu m) compared to the required electron beam resist thickness ($\approx$500 nm).
We use gold (Au) for the thin metal layer, as it is known to have a low contact resistance to graphene \cite{gahoi2016} and is unaffected by the HF acid required for the release of the actuator from the substrate.
However, Au has a low adhesion to the highly p-doped silicon.
Therefore, we use an additional thin metal layer of Cr or Tantalum (Ta) for enhancing the adhesion of the Au layer.

We implement the thin Au film as depicted in figure~\ref{fig4}(a).
A HF dip is performed to remove the natural oxide on the substrate.
This ensures a clean interface and thus increases the adhesion of the metal to the substrate.
Then an electron beam lithography step is used to deposit Ta~(7~nm) and Au~(30~nm) on an area that is slightly smaller than that of the Cr mask used for the CD actuator fabrication (see figure~\ref{fig4}(b)).
The Cr mask thus fully encapsulates the deposited metal (see figure~\ref{fig4}(c)) and thereby prevents any contamination of the DRIE chamber during etching.
Sometimes, we first only partially etch the highly p-doped silicon and then deposit an additional Cr layer for the fabrication of an electrostatic gate before etching completely through.
The method outlined here results in a thin layer of metal (see figure~\ref{fig4}(d)-(e)) over the entire actuator as shown in figure~\ref{fig4}(f), and results in a reduction of the device resistance to 236~$\Omega$~(black data points in figure~\ref{fig3}).

\begin{figure*}[hbt!]
	\makebox[\textwidth][c]{\includegraphics[width=\textwidth]{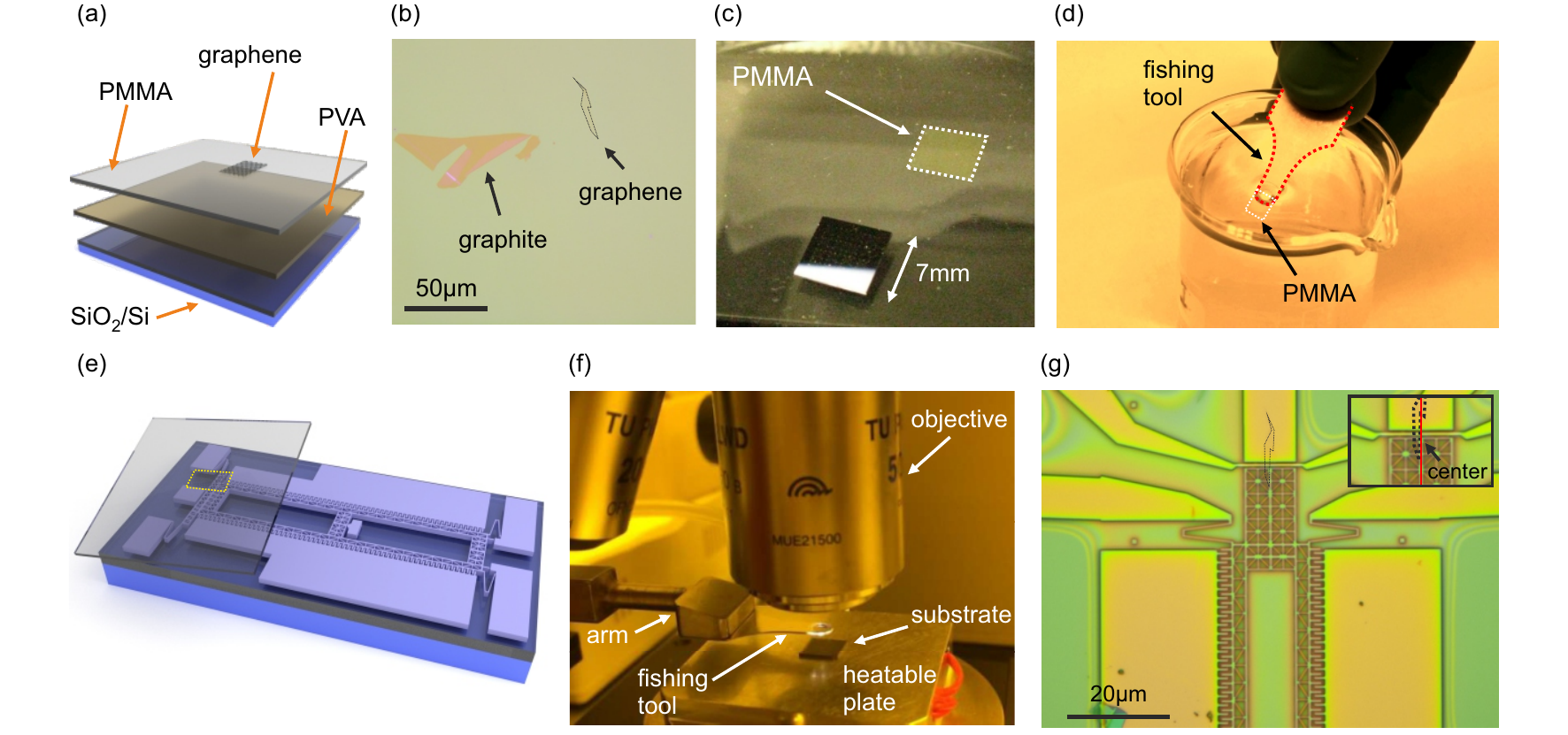}}
	\caption{Integration of graphene: (a) Mechanical exfoliation of a graphene onto a PVA/PMMA stack. (b) Identification of individual flakes with an optical microscope, (c) dissolving the PVA with DI water before picking it up with a fishing tool (d). (e) Schematic illustration of the transfer of the PMMA membrane with the graphene flake onto the actuator using the in setup (f). Panel (g) shows an optical image of the transferred graphene flake. The close-up depicts the misalignment ($\approx$3~\textmu m) between the graphene flake (see dashed line) and the center of the actuator (red line).}
	\vspace{-10pt}
    \label{fig5}
\end{figure*}
%
\subsection{Integration of graphene\label{S34}}
To integrate graphene onto the CD actuator, we use a method based on the one reported by Dean et al. \cite{dean2010boron}.
We spin coat a silicon substrate with a $\approx$300~nm thick layer of polyvinyl alcohol (PVA) and a $\approx$330~nm thick layer of PMMA (see figure~\ref{fig5}(a)).
These thicknesses are selected to optimize the contrast of individual graphene flakes in optical microscopy~\cite{casiraghi2007rayleigh}.
Then, we exfoliate graphene onto the prepared substrate.
Using an optical microscope, we identify an isolated graphene flake with typical dimensions of 2~\textmu m by 10~\textmu m (see figure~\ref{fig5}(b)).
After the identification of a suitable graphene flake, we submerge the substrate into deionized (DI) water to dissolve the PVA.
Note that the graphene does not get in contact with the water during this step.
Figure~\ref{fig5}(c) shows that the PMMA with an exfoliated graphene flake floats on top of the DI water.
This PMMA membrane is fished out of the DI water with an open circle that has a slightly smaller diameter than the membrane (see figure~\ref{fig5}(d)).
To transfer the PMMA membrane onto the actuator (see figures~\ref{fig5}(e-f)), we align the graphene flake with the transfer area of the actuator and make sure that they are in the same focus depth.
Then we use a nitrogen flow to press the PMMA membrane into contact with the actuator before heating up the actuator to 120~$^{\circ}$C to ensure good mechanical contact.
Finally, we lift up, i.e. mechanically remove, the fishing device leaving the graphene and the PMMA layer attached to the CD actuator (see figure~\ref{fig5}(g)).

\begin{figure}[!b]
	\includegraphics[width=86mm]{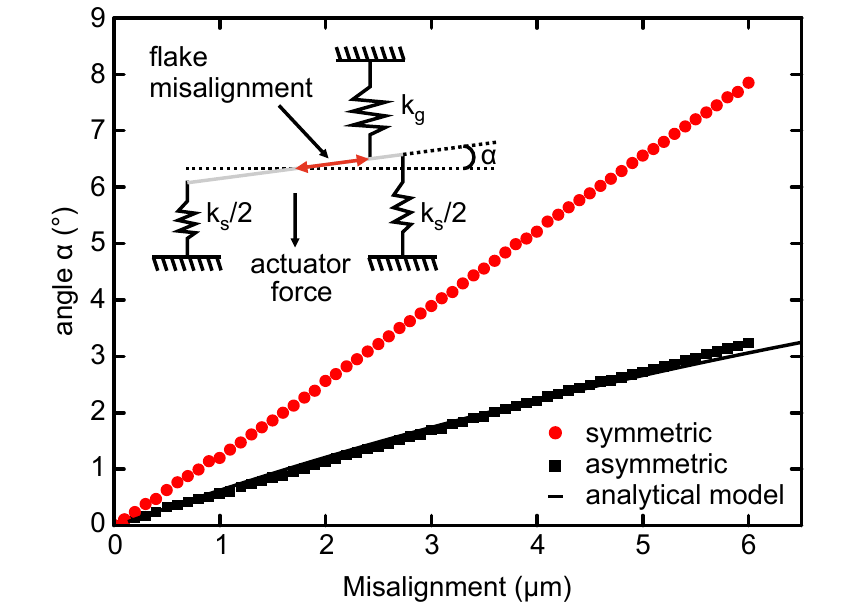}
	\caption{Finite element simulation of the rotation angle $\alpha$ when applying strain to a graphene flake as a function of the misalignment as defined by the dashed red line in the inset. The result for a fixed actuator force of 6.75~\textmu N indicates that $\alpha$ for the design with the asymmetrically spaced fingers depicted in Fig. \ref{fig1}b is up to 3 times smaller and therefore approaches the ideal actuator behavior with $\alpha = 0^{\circ}$. The simulation for the asymmetrically spaced fingers is in good agreement with the analytical solution for $\alpha$ when modeling the graphene as a spring $k_g = 540$~N/m connected to a single point on the CD actuator with spring constant $k_s = 6.5$~N/m, as depicted in the inset.}
	\label{fig6}
\end{figure}

\begin{figure*}[hbt!]
	\makebox[\textwidth][c]{\includegraphics[width=\textwidth]{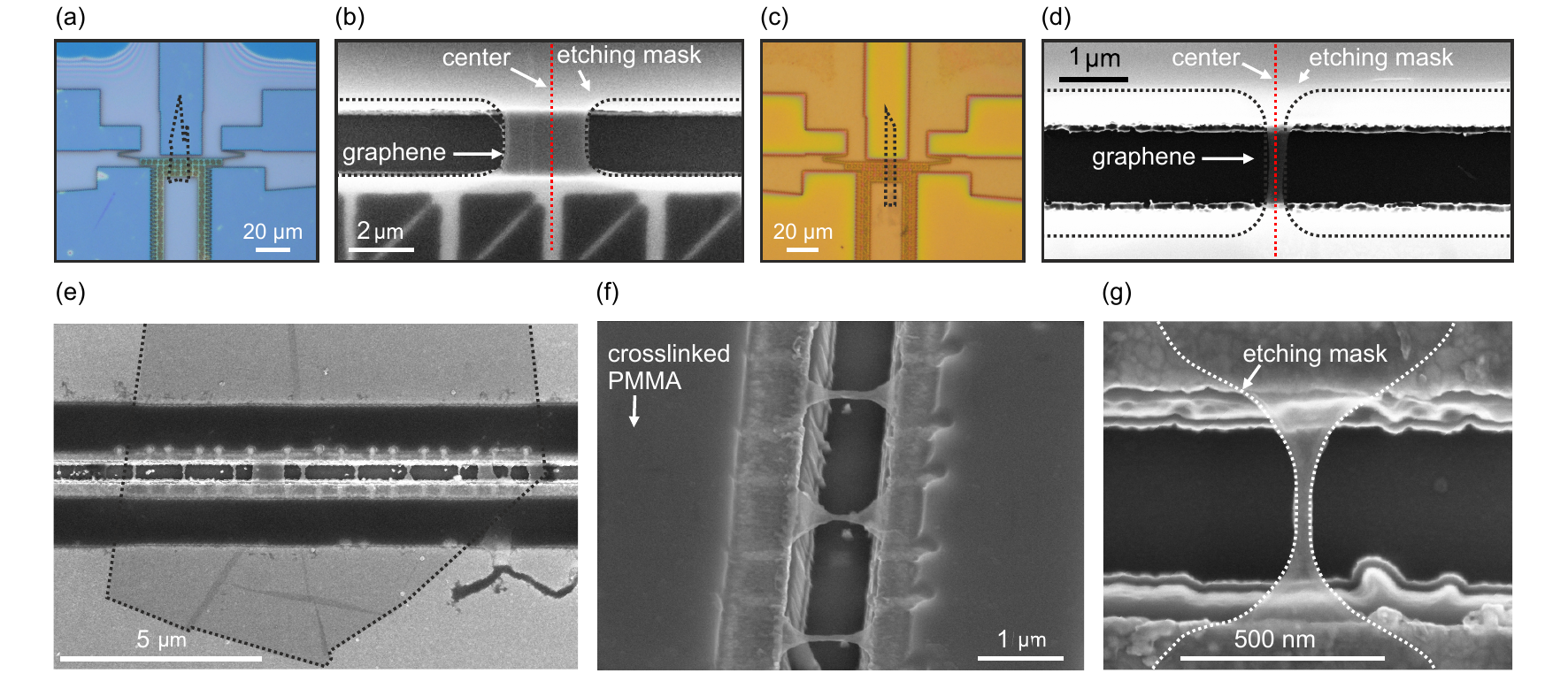}}
	\caption{(a-d) Structuring the graphene flakes after transfer onto an actuator: (a) and (c) Optical image of the flake prior structuring. (b) and (d) Scanning electron microscope image after structuring. Panels (b) and (d) show that the structured flake is almost perfectly aligned with the center of the actuator. The designed width of the graphene flakes was 2~\textmu m and 0.5~\textmu m in panels (b) and (d), respectively. Note the fringes indicated by the black arrows in panel (a), which indicate the height change of the simultaneously transferred PMMA membrane. (e-f) Structuring a suspended graphene flake over a 1~\textmu m wide trench: scanning electron microscope images of (e) the initial flake, (f) suspended nanoconstrictions, (g) and the narrowest constriction (30 nm) achieved with this technique.
	}\label{fig7}
	\vspace{-10pt}
\end{figure*}

\subsection{Patterning suspended graphene\label{S35}}
The integrated graphene flake is usually not exactly in the center of the transfer area of the actuator.
This misalignment, i.e. the distance to the center of the transfer area, can be up to 5~\textmu m, which affects the operation of the actuator.
The graphene flake acts as an additional spring $k_\text{g}$ connected to the actuator.
Consequently, the force of the actuator on the transfer area will be unevenly distributed (see inset figure~\ref{fig6}).
This results in a rotation of the actuator.

We quantify the rotation of the actuator for the designs depicted in figure~\ref{fig1}(a) and figure~\ref{fig1}(c) by extracting the rotation angle $\alpha$ as a function of the misalignment for a fixed force of 6.75~\textmu N from a finite element simulation with COMSOL \cite{comsol}.
In the simulation, the spring constant $k_\text{g}$ of the graphene was 540~N/m and the spring constant $k_\text{s}$ of the CD actuator was 6.5~N/m.
The spring constant $k_\text{g} = Y_\text{2D} W/L$ was chosen to approximate the spring constant of a typically transferred graphene flake with a width $W$ of $\approx3$~\textmu m and a suspended length $L$ of 2~\textmu m. Here, $Y_\text{2D} = 340$~N/m is the two-dimensional Young's modulus of graphene \cite{lee2008}.
The result, shown in figure~\ref{fig6}, illustrates that our CD design has up to 3 times less rotation compared to the symmetric actuator design.
In addition, the simulation for the design depicted in figure~\ref{fig1}(a) is in good agreement with the analytical model for $\alpha$ (inset figure~\ref{fig6}).
Nevertheless, the rotation angle $\alpha$ generates a non-negligible shear strain in the integrated graphene sheet.

A viable method to circumvent the misalignment problem is to transfer graphene flakes that are comparable in size to the transfer area of the CD actuator (see figure~\ref{fig1}(a)).
We then pattern the simultaneously transferred PMMA membrane with an electron dose of $\approx120$~\textmu C/cm$^\text{2}$ at 20~kV.
After development of the PMMA membrane, we structure the suspended graphene with reactive ion etching using SF$_6$ with a RF power of 60~W for 6~s and thereby regain control over the position and geometry of the graphene flake.
Figure~\ref{fig7}(a) shows an optical image of a transferred PMMA layer with exfoliated graphene.
Note the fringes indicated by the black arrows in figure~\ref{fig7}(a), which indicate the height change of the PMMA membrane when going from the CD actuator to the substrate.
By using reactive ion etching, we are able to structure the graphene such that it is well aligned with the center of the transfer area (see SEM image in figure~\ref{fig7}(b)).
A second example is depicted in figures~\ref{fig7}(c-d), in which we even reduced the graphene width to 500~nm.
The described patterning process gives us control over the geometry of suspended graphene flakes and is not only applicable to our CD devices.
To show this, we prepared a substrate with a 0.3~\textmu m thick highly p-doped silicon layer (see Sect.~3.1) and used DRIE to structure a 1~\textmu m wide and 350~nm deep trench in it (see Sect.~3.2).
After transferring a PMMA layer with exfoliated graphene over the trench (see figure~\ref{fig7}(e)), we structured several suspended nanoconstrictions (see tilted SEM image in figure~\ref{fig7}(f)).
Figure~\ref{fig7}(g) shows a zoom-in of the narrowest (30~nm) constriction we made.
By this technique, we are thus able to build suspended structures with a minimum feature size up to 30~nm.
This promising approach allows for suspended nanoconstriction devices (see figures~\ref{fig7}(e-g)) and potentially quantum dot devices \cite{terres2016size}.
Finally, we point out that this method of structuring a transferred graphene flake is generally applicable to other suspended graphene devices such as resonators as well as to two-dimensional materials in general.

After structuring the suspended graphene flake, we have to clamp it to the actuator.
Hereby, we prevent the removal of the graphene flake during the release of the device from the substrate.
In addition, strong clamping of the graphene to the actuator is required to induce strain in the graphene as well as to prevent slipping of the graphene.
Ideally, the graphene flake ruptures before the clamping does.
We found that clamping the graphene by cross-linking the remaining PMMA after structuring that is on top of both the actuator and the graphene sheet with an electron dose of 10.000~\textmu C/cm$^2$ fulfils all the mentioned requirements \cite{goldsche2018}.

\subsection{Release from the substrate\label{S36}}
After the transfer and patterning of the graphene, we dissolve the non-crosslinked PMMA in acetone and transfer the actuator into DI water.
We finally release the combs that are connected to a grated structure by submerging the entire device into a 10\% HF acid solution.
The grated structure ensures that the HF solution reaches all the SiO$_2$ underneath.
Consequently, this comb is suspended and freely movable whereas the other comb remains fixed to the substrate.
The HF solution has an etch rate of $\approx$1~nm/s, which sets the total etching time to $\approx$15~minutes.
After etching, we flush the device with DI water for 10~min.
Then, we shortly flush the device with isopropanol.
As the surface tension is strong enough to pull the actuator down to the substrate, we finally use a critical point dryer (CPD) to remove the liquids from the actuator.

\begin{figure*}[hbt!]
	\makebox[\textwidth][c]{\includegraphics[width=\textwidth]{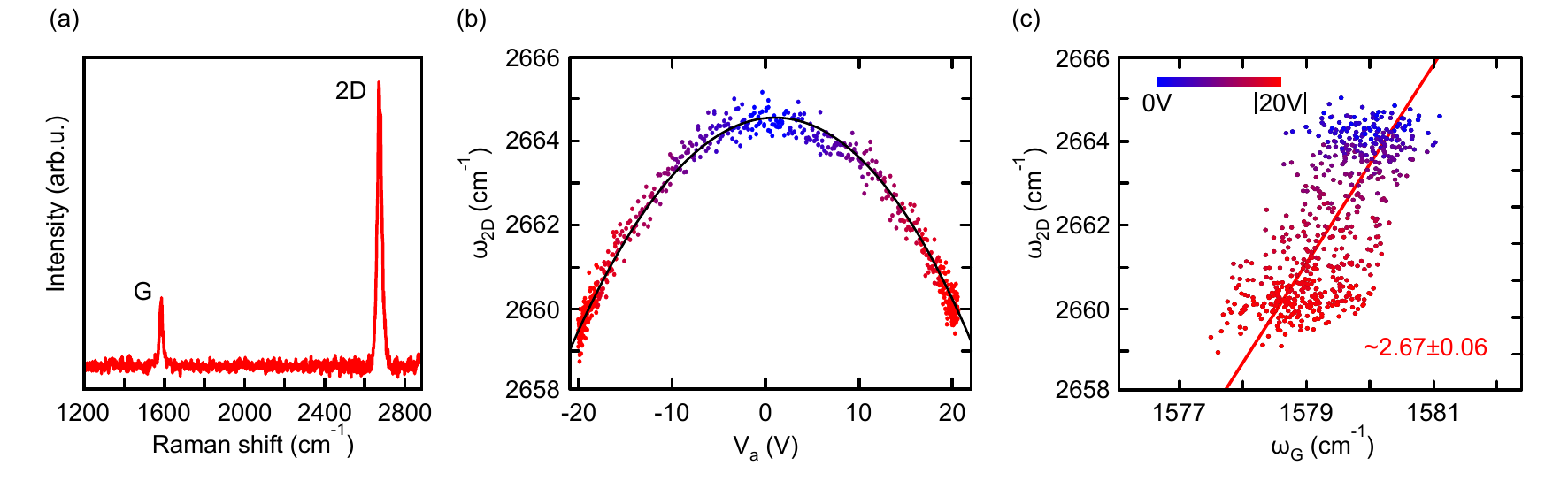}}
	\caption{(a) A typical Raman spectrum of graphene obtained after transfer onto the CD actuator. (b) shift of the Raman active 2D-mode as a function of applied actuator potential. The parabolic fit (black line) is in agreement with the electrostatic force $\sim V_\text{a}^2$. (c) the relative shift of the Raman active 2D- and G-mode have an overall linear correlation with a slope of~2.67, which is a clear indication that the actuator induces strain into the graphene sheet. We attribute the relatively large spread to spatial inhomogeneities in the doping of the graphene sheet within the laser spot size \cite{lee2012optical}.
	}\label{fig8}
	\vspace{-10pt}
\end{figure*}
\begin{figure*}[!hbt]
	\includegraphics[width=166.4mm]{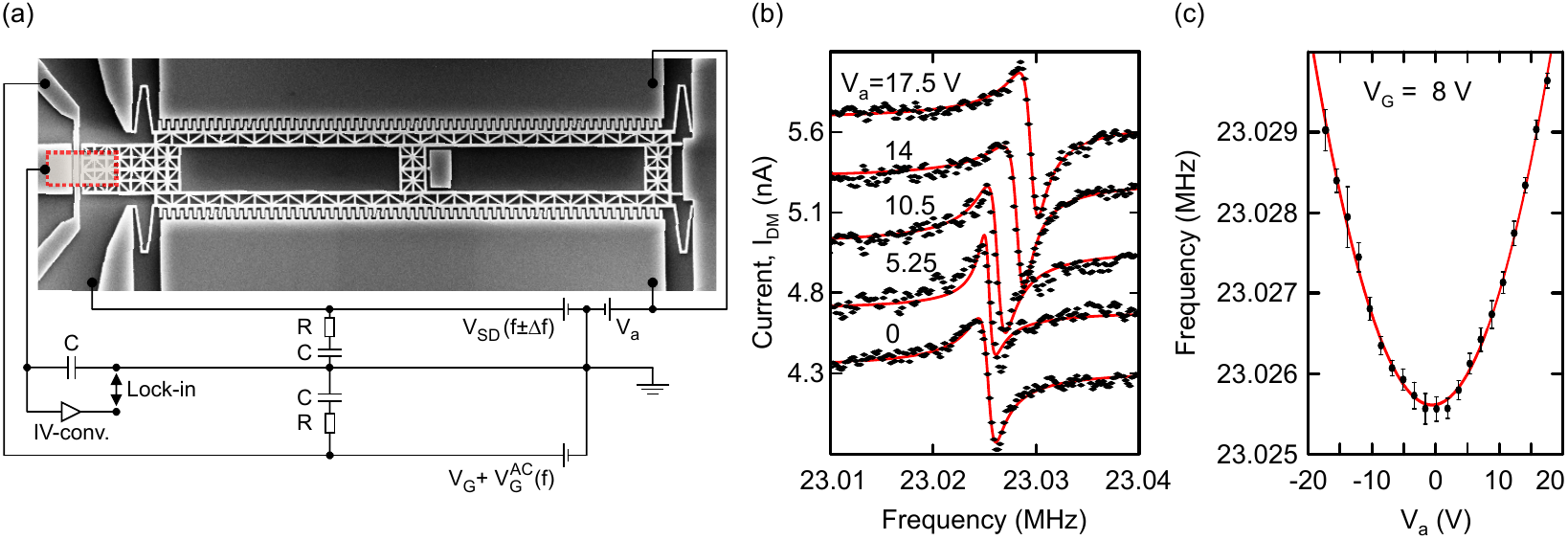}
	\caption{(a) A SEM image of the device with its electrical connections. An AC signal ($V_\text{SD}(f\pm\Delta f)$) at frequency $f\pm\Delta f$ biases the graphene resonator (red dashed box) while a sum of DC and AC voltages ($V_\text{G} + V_\text{G}^\text{AC}(f)$) are applied to the gate underneath. The down-mixed current $I_\text{DM}$ at frequency $\Delta f$ is measured with an IV-converter and a lock-in amplifier. A DC voltage $V_\text{a}$ controls the actuator. The capacitors $C = 100$ nF and resistors $R = 50$ $\Omega$ are chosen for impedance matching and decoupling any high frequency signals. (b) The measured down-mixing current $I_\text{DM}$ as a function of the actuation frequency $f$ for different $V_\text{a}$ shows a resonance around 23.0254~MHz. The curves are offset for clarity. The red lines represent fits to a non-zero phase Lorentzian \cite{chen2009performance,zande2010large}. (c) The extracted resonance frequency as a function of applied $V_\text{a}$. The parabolic fit (red line) is in agreement with the electrostatic force $\sim V_\text{a}^2$ that increases the strain in the resonator.
	}\label{fig9}
	\vspace{-10pt}
\end{figure*}

\section{Characterization\label{S4}}

To show the functionality of our CD actuators with integrated graphene, we present measurements performed on two different devices.
In the first set of measurements, depicted in figure~\ref{fig8}, we performed confocal Raman spectroscopy on a controllably strained graphene flake.
We use linearly polarized laser light ($\lambda\approx532$~nm) with a power of 0.5~mW and a spot size of $\approx500$~nm.
The scattered light is detected by a CCD spectrometer with a grating of 1200 lines/mm.
With Raman spectroscopy one probes optical phonons via the inelastic scattering of light.
Note that the frequency of optical phonons depends on the crystal lattice and therefore on the applied strain.
The Raman spectrum of graphene contains a peak at $\omega_\text{G}\sim 1580$ cm$^\text{-1}$ and at $\omega_\text{2D}\sim 2665$ cm$^\text{-1}$ (see figure~\ref{fig8}(a)), which are the so-called G- and 2D-mode peaks, respectively.
As the electrostatic force of the CD actuator on the graphene flake is proportional to $V_\text{a}^2$, the induced strain in the graphene sheet should show the same dependence.
As figure~\ref{fig8}(b) shows, we indeed observe a parabolic tuning $V_\text{a}^2$ of $\omega_\text{2D}$.
The G-peak frequency shows similar behavior (not shown).
Combining the measured $\omega_\text{G}$ and $\omega_\text{2D}$ for different $V_\text{a}$ into figure~\ref{fig8}(c), we observe that the data points follow roughly a straight line with a slope of 2.67.
We attribute the relatively large spread to spatial inhomogeneities in the doping of the graphene sheet within the laser spot size \cite{lee2012optical}.
Experiments reported in the literature show that this slope is characteristic for strain induced frequency shifts \cite{goldsche2018,lee2012optical,mohiuddin2009uniaxial,yoon2011strain}.
By using the reported value of -83 cm$^\text{-1}/$\% for the frequency shift of $\omega_\text{2D}$ per unit of applied uniaxial strain \cite{mohiuddin2009uniaxial}, we find a maximum strain value of 0.06\%.
Note that the spring constant of the graphene is much stiffer than that of the actuator such that 20~V is needed to elongate the graphene by only 1.2~nm (c.f. Fig.~\ref{fig1}).
A detailed discussion on the effect of the spring constants in relation to the measured Raman shifts can be found in Ref.~\onlinecite{goldsche2018}.
So far, we reached a maximum strain value of 1.2\% in a controllable and reproducible fashion with the CD actuator presented in this work \cite{goldsche2018}.
This shows that the actuator is strong enough to induce a strain value above 1\% in suspended graphene flakes.

In a second test experiment, we measure a mechanical resonance frequency of a graphene flake integrated onto an CD actuator operating at 2.2~K using an amplitude modulated down-mixing scheme (see figure~\ref{fig9}(a)) \cite{verbiest2016,chen2009performance}.
In this scheme, we apply a DC voltage $V_\text{G}$ as well as an AC voltage $V_\text{G}^\text{AC}$ at frequency $f$ to the electrostatic gate below the graphene in order to tune the charge carrier density and to mechanically actuate it.
In addition, a small modulation voltage $V_\text{SD} = 10$~mV at frequency $f \pm \Delta f$ is applied across the graphene membrane to generate a current $I_\text{DM}$ at frequency $\Delta f$, which is amplified with an IV-converter and measured with an UHF lock-in amplifier from Z\"{u}rich Instruments.
Figure~\ref{fig9}(b) shows the unprocessed $I_\text{DM}$ around the resonance peak for different $V_\text{a}$ and the corresponding fit with a non-zero phase Lorentzian (red lines) \cite{chen2009performance,zande2010large}.
The extracted resonance frequency and quality factor are 23.0254~MHz and 15.220, respectively.
These numbers are comparable with similar experiments performed on graphene resonators reported in the literature \cite{chen2009performance,chen2013graphene}.
In analogy to the Raman measurement, the electrostatic force generated by applying a voltage $V_\text{a}$ to the actuator strains the graphene sheet.
The induced strain and thus tension in the graphene sheet naturally results in an increase of the resonance frequency, as depicted in figure~\ref{fig9}(c).
Note that the applied $V_\text{SD}$ also results in a constant Joule heating of 7~nW, which is independent from $V_\text{a}$ and therefore cannot explain the observed change in resonance frequency as observed in Ref.~\onlinecite{ye2018}.
The observed parabolic tuning of the resonance frequency (red line in figure~\ref{fig9}(c)) is in agreement with the applied electrostatic force.
Moreover, this measurement proofs the functionality of the CD actuators at cryogenic temperatures.

\section{Conclusion and Outlook\label{S5}}

We integrated suspended graphene on CD actuators to perform strain-dependent measurements.
We presented an optimized CD actuator design and described in detail the fabrication process.
We presented a DRIE process to fabricate CD actuators with a scallop depth below 5~nm and a process to implement a gold layer on top of the Si actuator.
In particular, the actuator resistance at room temperature reduces from 51.6~k$\Omega$ to 236~$\Omega$ after implementing the gold.
This enables sensitive electrical measurements as a function of strain on integrated suspended graphene.
The functionality of the fabricated actuators was shown by measuring the Raman spectrum and the mechanical resonance frequency of an integrated graphene sheet as a function of induced strain.

Moreover, we introduced a method to structure suspended graphene flakes.
When applying this method to integrated graphene flakes on actuators, ideal device functionality can be realized.
The minimum feature size achievable with this method is up to 30~nm, which enables device concepts such as suspended quantum dots and nanoconstrictions.
This method is also applicable to other suspended graphene devices such as resonators.

The methods for integrating graphene and structuring suspended graphene presented in this work are generally applicable to other two-dimensional materials.
Therefore, this work provides a toolbox for strain dependent (transport) measurements in two-dimensional materials, which is not only interesting from the fundamental point of view but also for the development of novel device concepts for applications in the fields of sensors and transducers.

\section{Acknowledgement}

The authors thank S. Trellenkamp and B. Hermanns for the support and discussions concerning the actuator fabrication.
We acknowledge support from the Helmholtz Nano Facility (HNF) \cite{hnf2017} and funding from the ERC (GA-Nr. 280140).


\end{document}